\documentclass[aps,prl,twocolumn,superscriptaddress,longbibliography]{revtex4}
\usepackage{graphicx}
\usepackage{amssymb,amsmath}
\usepackage{bm}
\usepackage{dcolumn}
\usepackage{subfigure}
\usepackage{float}
\usepackage[OT1]{fontenc} 
\usepackage{url}
\usepackage{slashed}
\usepackage{color}
\usepackage{verbatim}
\usepackage{enumitem}
\usepackage{txfonts}
\usepackage{soul,physics}
\usepackage[driverfallback=dvipdfm]{hyperref}
\hypersetup{pdfpagemode=FullScreen,colorlinks=true,breaklinks,urlcolor=blue,linkcolor=blue,citecolor=blue}
\usepackage{natbib}

\usepackage{subfigure}
\usepackage{comment}
\usepackage{hyperref}
\usepackage{amssymb}
\usepackage{bm}
\usepackage{graphicx}
\usepackage{amsmath,amssymb}
\usepackage{tikz,fp}
\usepackage{tikz-cd}
\usetikzlibrary{arrows}
\usetikzlibrary{intersections}
\usetikzlibrary{shapes.geometric}
\usetikzlibrary{decorations.pathmorphing, patterns,shapes,fixedpointarithmetic}
\usetikzlibrary{decorations.markings}

\pgfdeclarepatternformonly{south west lines}{\pgfqpoint{-0pt}{-0pt}}{\pgfqpoint{3pt}{3pt}}{\pgfqpoint{3pt}{3pt}}{
	\pgfsetlinewidth{0.4pt}
	\pgfpathmoveto{\pgfqpoint{0pt}{0pt}}
	\pgfpathlineto{\pgfqpoint{3pt}{3pt}}
	\pgfpathmoveto{\pgfqpoint{2.8pt}{-.2pt}}
	\pgfpathlineto{\pgfqpoint{3.2pt}{.2pt}}
	\pgfpathmoveto{\pgfqpoint{-.2pt}{2.8pt}}
	\pgfpathlineto{\pgfqpoint{.2pt}{3.2pt}}
	\pgfusepath{stroke}}

\tikzset{
	mid arrow/.style={postaction={decorate,decoration={
				markings,
				mark=at position .575 with {\arrow{stealth}}
	}}},
	near arrow/.style={postaction={decorate,decoration={
				markings,
				mark=at position .275 with {\arrow{stealth}}
	}}},
	far arrow/.style={postaction={decorate,decoration={
				markings,
				mark=at position .800 with {\arrow{stealth}}
	}}},
	snake arrow/.style={fixed point arithmetic, decorate, decoration={snake,amplitude=2pt, segment length=11pt},postaction={decoration={markings,mark=at position 0.625 with {\arrow{stealth}}},decorate}},
  my snake/.style={fixed point arithmetic, decorate, decoration={snake,amplitude=2pt, segment length=11pt}}
}
\tikzset{
  baseline = -0.5ex,
  wavy/.style = {
    thick,
    decorate,
    decoration={snake,amplitude=2pt,segment length=5pt}},
  sdot/.style = {
    circle,
    draw=none,
    fill=black,
    minimum size=2.5pt,
    inner sep=0pt},
  bdot/.style = {
    circle,
    draw=none,
    fill=black,
    minimum size=4pt,
    inner sep=0pt},
  svertex/.style = {
    circle,
    draw=black,
    thick,
    fill=lightgray,
    minimum size=8pt,
    inner sep=1pt},
  bvertex/.style = {
    circle,
    draw=black,
    thick,
    fill=lightgray,
    minimum size=24pt},
  bvertexsmall/.style = {
    circle,
    draw=black,
    thick,
    fill=lightgray,
    minimum size=7pt},
  bvertexnormal/.style = {
    circle,
    draw=black,
    thick,
    fill=lightgray,
    minimum size=14pt},
  dvertex/.style = {
    circle,
    draw=black,
    thick,
    fill=gray,
    minimum size=25pt}}

\begin{document}
	
	\title{Environment-Induced Information Scrambling Transition with Charge Conservations}
	
	\author{Pengfei Zhang}
  \affiliation{Department of Physics, Fudan University, Shanghai, 200438, China}
    \affiliation{Center for Field Theory and Particle Physics, Fudan University, Shanghai, 200438, China}
    \affiliation{Shanghai Qi Zhi Institute, AI Tower, Xuhui District, Shanghai 200232, China}
	\author{Zhenhua Yu}
	\thanks{huazhenyu2000@gmail.com}
	\affiliation{Guangdong Provincial Key Laboratory of Quantum Metrology and Sensing, School of Physics and Astronomy, Sun Yat-Sen University (Zhuhai Campus), Zhuhai 519082, China}
  \affiliation{State Key Laboratory of Optoelectronic Materials and Technologies, Sun Yat-Sen University (Guangzhou Campus), Guangzhou 510275, China}
	\date{\today}

	\begin{abstract}
	In generic closed quantum systems, the complexity of operators increases under time evolution governed by the Heisenberg equation, reflecting the scrambling of local quantum information. However, when systems interact with an external environment, the system-environment coupling allows operators to escape from the system, inducing a dynamical transition between the scrambling phase and the dissipative phase. This transition is known as the environment-induced information scrambling transition, originally proposed in Majorana fermion systems. In this work, we advance this dicovery by investigating the transition in charge-conserved systems with space-time randomness. We construct solvable  Brownian Sachdev-Ye-Kitaev models of complex fermions coupled to an environment, enabling the analytical computation of operator growth. We determine the critical dissipation strength, which is proportional to $n(1-n)$ with $n$ being the density of the complex fermions, arising from the suppression in the quantum Lyapunov exponent due to the Pauli blockade in the scattering process. We further analyze the density dependence of maximally scrambled operators at late time. Our results shed light on the intriguing interplay between information scrambling, dissipation, and conservation laws.
	\end{abstract}
	
	\maketitle

  \section{ Introduction} 
  Recent years have witnessed breakthroughs in understanding the dynamics of quantum information in many-body systems. In particular, there has been considerable attention focusing on information scrambling \cite{Hayden:2007cs,Sekino:2008he,Shenker:2014cwa,Roberts:2014isa,2015Natur.528...77I,Li:2016xhw,Garttner:2016mqj,2019Sci...364..260B,2019arXiv190206628S,2019Natur.567...61L,2020PhRvL.124x0505J,Blok:2020may,2021PhRvA.104a2402D,2021PhRvA.104f2406D,Mi:2021gdf,Cotler:2022fin,2022PhRvA.105e2232S}, which describes how local information disperses throughout the entire system. As an illustration, suppose we initially encode a single-qubit state $|\psi\rangle$ locally in a many-body wavefunction by preparing a product state $|\Psi\rangle=|\psi\rangle \otimes |0\rangle \otimes...\otimes |0\rangle$. 
  %The manipulation of this information corresponds to the application of operators on the first site $\{\hat O_1\}$. 
  We then proceed to evolve the entire system in time, described by an evolution operator $\hat{U}$. Consequently, the state of the system becomes $|\Psi(t)\rangle =\hat{U}~|\Psi\rangle$. In general, this evolution will build up entanglement between different qubits in the many-body system, causing the encoded information to become extended. To read off the encoded information by only observing operators $\{\hat O^m_1\}$ acting locally on the first qubit, one should apply to $|\Psi(t)\rangle$ the backwardly evolved operators $\{\hat O_1^m(-t)\equiv \hat U\hat O^m_1\hat{U}^\dagger\} $, which is highly complex and non-local. A quantitative measure of operator complexity is provided by the operator size distribution, extensively investigated in solvable models or through numerical simulations \cite{Roberts:2014isa,Nahum:2017yvy,Qi:2018bje,vonKeyserlingk:2017dyr,Khemani:2017nda,Hunter-Jones:2018otn,Chen:2019klo,PhysRevLett.122.216601,Chen:2020bmq, Lucas:2020pgj,Yin:2020oze,Zhou:2021syv,Dias:2021ncd,2021PhRvR...3c2057W,gu2022two,Zhang:2022fma,Liu:2023lyu,Gao:2023kwk,Gao:2024jtb}. For systems with all-to-all interactions, the operator size exhibits exponential growth in the early-time regime and eventually "thermalizes" to a maximally scrambled operator modeled by a uniform distribution among all operators.

      \begin{figure}[tb]
    \centering
    \includegraphics[width=0.8\linewidth]{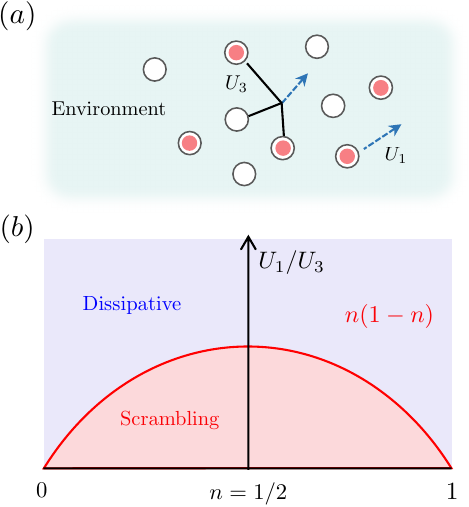}
    \caption{(a) Schematics of the all-to-all interacting systems with charge conservation. Here, each red dot corresponds to a charge occupation. $U_3$ corresponds to an interaction between three system fermions and a single environment fermion, and $U_1$ represents the hopping between the system and environment. (b) The phase diagram of the model, in which the information scrambling transition occurs at $U_1/U_3 = n(1-n)$.}
    \label{fig:schemticas}
  \end{figure}

  Later, the study of information scrambling has been extended to systems that are embedded in environments \cite{Zhang:2023xrr,Bhattacharya:2022gbz,2022arXiv220713603L,schuster2022operator,Weinstein:2022yce}. In the presence of system-environment couplings, an operator may disappear from the system and disperses into the environment. This process can be viewed as a dissipation in the operator space. Refs. \cite{Zhang:2023xrr,Weinstein:2022yce} propose that actually an environment-induced information scrambling transition can occur when the system-environment couplings are fine tuned; this transition separates a scrambling phase and a dissipative phase. In the scrambling phase, the operator size within the system continues to increase to an order of $N$ (where $N$ is the system size), as in closed systems. Conversely, in the dissipative phase, the size decreases exponentially over time. Additionally, it is noted that the distribution of operator sizes displays a power-law decay at the critical point. Explicit examples include random circuit models \cite{Weinstein:2022yce} and Brownian SYK models \cite{kitaev2015simple,maldacena2016remarks,kitaev2018soft,Saad:2018bqo,Sunderhauf:2019djv} with Majorana fermions \cite{Zhang:2023xrr}. Both models feature space-time randomness and lacking any discernible global symmetry.

However, it is well-established that conservation laws exert profound implications on the dynamics of quantum many-body systems. For instance, in systems with charge conservation, the out-of-time-order correlator \cite{Shenker:2014cwa,Roberts:2014isa,1969JETP...28.1200L,kitaev2014hidden} in random circuits exhibits power-law tails attributed to the diffusion of local charges \cite{PhysRevX.8.031058,PhysRevX.8.031057}. It is also known that the Rényi entropies (with Rényi index $\alpha>1$) of a half-infinite system grow as $\leq\sqrt{t\ln t}$ \cite{2019arXiv190200977H,2019PhRvL.122y0602R} after a quantum quench. In systems at low/high densities $n$, both the quantum Lyapunov exponent and the butterfly velocity are suppressed by a polynomial of $n$ due to the Pauli blockade of the effective Hilbert space \cite{10.21468/SciPostPhys.9.5.071,Zhang:2023vpm,Agarwal:2020yky}. Therefore, it is interesting to explore the the fate of the environment-induced information scrambling transition in systems with charge conservations.

  In this work, we construct a solvable complex Brownian SYK model embedded in an environment, as illustrated in FIG. \ref{fig:schemticas} (a). The total system exhibits a weak U(1) symmetry, allowing for the interchange of charges between the system and the environment. By analyzing generic system-environment couplings, we identify dominant terms in the low/high-density limit. The main contribution to the information scrambling in the system comes from a four-body interaction, yielding a quantum Lyapunov exponent $\varkappa_{\text{eff}}\propto n(1-n) U_3$, where $U_3$ is the interaction strength. For comparison, dissipation is governed by a direct hopping between the system and the environment, with an effective dissipation rate $U_1$ independent of the density $n$. Consequently, we expect that the information scrambling transition shall occur at $U_1/U_3\propto n(1-n)$, as sketched in FIG. \ref{fig:schemticas} (b). This expectation agrees with an analytically calculation using the complex Brownian SYK model. We further analyze the late-time regime using the scramblon effective theory \cite{gu2022two,Stanford:2021bhl}, which unveils the renormalization of maximally scrambled operator.  

  \section{Model and Two-point functions}
  We consider a Brownian SYK model of $N$ complex fermions $\hat{c}_j$. The Hamiltonian of the system reads
  \begin{equation}
  \hat{H}_S=\sum_q\sum_{i_1<...<i_{\frac{q}{2}}}\sum_{j_1<...<j_{\frac{q}{2}}}J_{\{i_l,j_l\}}(t)\hat{c}^\dagger_{i_1}\hat{c}^\dagger_{i_2}...\hat{c}^\dagger_{i_{\frac{q}{2}}}\hat{c}^{}_{j_1}\hat{c}^{}_{j_2}...\hat{c}^{}_{j_{\frac{q}{2}}},
  \end{equation}
  where the coupling strengthes $J_{\{i_l,j_l\}}(t)$ are independent Brownian variables with zero expectation and 
  \begin{equation}
  \overline{J_{\{i_l,j_l\}}(t) J_{\{i_l,j_l\}}(t')^*}=\frac{J_q\delta(t-t')}{\frac{q}{2}!(\frac{q}{2}-1)!N^{q-1}}.
  \end{equation}
  Here, the normalization factor is chosen to cancel the symmetry factor of self-energies. We then couple the system to a microscopic environment, which contains complex fermions $\hat{e}_a$ with $a\in\{1,2,...,M\}$ \cite{Chen:2017dbb,Zhang:2019fcy,Almheiri:2019jqq}. We assume $M\gg N$ to ensure that the environment contains much more degrees of freedoms. We consider a generic coupling Hamiltonian between the system and the environment
  \begin{equation}
  \begin{aligned}
  \hat H_{SE}={\sum_{\{p_l\}}}'\sum_{\{i_l,j_l,a_k,b_k\}}&U_{\{i_l,j_l,a_k,b_k\}}(t)\hat{c}^\dagger_{i_1}...\hat{c}^\dagger_{i_{p_1}}\hat{c}_{j_1}...\hat{c}_{j_{p_2}}\\
  &\times\hat{e}^\dagger_{a_1}...\hat{e}^\dagger_{a_{p_3}}\hat{e}_{b_1}...\hat{e}_{b_{p_4}},\label{hse}
  \end{aligned}
  \end{equation}
  where we use $(p_1,p_2,p_3,p_4)$ to count the number of $(\hat{c}^\dagger,\hat{c},\hat{e}^\dagger,\hat{e})$ operators. We impose the conservation of the particle number $\hat{Q}=\sum_i \hat{c}^\dagger_i \hat{c}_i+\sum_a \hat{e}^\dagger_a \hat{e}_a$ by restricting the summation over $\{p_l\}$ to partitions satisfying $p_1+p_3=p_2+p_4$. The variance of the system-environment couplings are
  \begin{equation}
  \overline{U_{\{i_l,j_l,a_k,b_k\}}(t)U_{\{i_l,j_l,a_k,b_k\}}(t')^*}=\frac{p_2 U_{\{p_l\}}\delta(t-t')}{\prod_{l=1}^4(p_l!)~N^{p_1+p_2-1}M^{p_3+p_4}}.
  \end{equation}

  Since the time dependence of the coupling strengths breaks time translational invariance, the only tuning parameter for the steady state of the model is the charge density $n$. Employing a grand canonical ensemble, the steady-state density matrix is given by $\rho=\mathcal{Z}^{-1}e^{-\mu \hat{Q}}$, where $\mathcal{Z}=\text{tr}[e^{-\mu \hat{Q}}]$. The density $n$ can be related to the chemical potential $\mu$ by
  \begin{equation}
  n=\langle \hat{c}^\dagger_i \hat{c}_i \rangle=\langle\hat{e}^\dagger_a \hat{e}_a \rangle=\frac{1}{e^{\mu}+1}.
  \end{equation}
  The steady-state properties can be analyzed using a Keldysh path-integral approach, which includes a forward evolution branch $u$ and a backward evolution branch $d$: 
  \begin{equation}\label{eqn:singlecontoursketch}
\begin{tikzpicture}[scale=0.65,baseline={([yshift=0pt]current bounding box.center)}]

\draw[thick,blue,far arrow] (0pt,0pt)--(140pt,0pt);
\draw[thick,blue,far arrow] (140pt,-4pt)--(0pt,-4pt);
\filldraw (140pt,-2pt) circle (2pt) node[above right]{\scriptsize $\infty$};
\node at (70pt,10pt) {\scriptsize $u$};
\node at (70pt,-14pt) {\scriptsize $d$};
\filldraw (0pt,-2pt) circle (0pt) node[above left]{\scriptsize $-\infty$};

\draw[thick,blue] (0pt,0pt)--(0pt,10pt);
\draw[thick,blue] (-2pt,7pt)--(2pt,7pt);
\draw[thick,blue] (0pt,-4pt)--(0pt,-30pt);
\draw[thick,blue] (-2pt,-27pt)--(2pt,-27pt);

\node[svertex] at (0pt,-16pt) {\scriptsize$\rho$};

\end{tikzpicture}
\end{equation}
The two-point function on this contour is a two-by-two matrix $G^{ss'}(t)=\langle c^s_{i}(t)\bar{c}^{s'}_{i}(0)\rangle$, where $s,s'\in\{u,d\}$ labels fields on different branches. In the large-$N$ limit, the self-energy only receives contributions from melon diagrams; the Schwinger-Dyson equation reads \cite{Zhang:2020jhn}
 \begin{equation}
 \begin{aligned}
 &
  \begin{pmatrix}
 G^{uu} & G^{ud}\\
 G^{du} & G^{dd}
 \end{pmatrix}=\begin{pmatrix}
 -i\omega-\Sigma^{uu} & -\Sigma^{ud}\\
 -\Sigma^{du} & i\omega-\Sigma^{dd}
 \end{pmatrix}^{-1},
\end{aligned}
 \end{equation}
 with self-energies
 \begin{equation}
 \begin{aligned}
&\Sigma^{uu}(\omega)=\Sigma^{dd}(\omega)=-\frac{\Gamma}{2}(1-2n),\\
&\Sigma^{ud}(\omega)=-\Gamma n,\ \ \ \ \ \ \ 
\Sigma^{du}(\omega)=\Gamma (1-n),
\end{aligned}
 \end{equation}
 where we have introduced the quasi-particle lifetime 
 \begin{equation}
\Gamma=\sum_qJ_q [n(1-n)]^{\frac{q}{2}-1}+{\sum_{\{p_l\}}}' U_{\{p_l\}} [n(1-n)]^{\frac{\sum_l p_l}{2}-1}.
 \end{equation}
 Performing the inverse Fourier transform, we find Green's functions in the time domain
 \begin{equation}\label{eqn:BrownianG}
G(t)=\begin{pmatrix}
\frac{1}{2}-n+\frac{1}{2}\text{sgn}(t)&-n\\
1-n& \frac{1}{2}-n-\frac{1}{2}\text{sgn}(t)
\end{pmatrix}e^{-\frac{\Gamma |t|}{2}}.
 \end{equation}
 This determines the retarded and advanced Green's functions 
 \begin{equation}
G^\text{R/A}(t)=\mp i\theta(\pm t)~\text{tr}\left[\rho \{c_i^{}(t),c_i^{\dagger}(0)\}\right]=\mp i\theta(\pm t)e^{-\frac{\Gamma |t|}{2}}.
 \end{equation}

  \section{Operator size for complex fermions}
  We investigate the information scrambling in our model by computing the operator size growth. Our aim is to study operator growth in systems embedded in an environment at arbitrary density $n$. To motivate its general definition, we first consider the half-filling case $\mu=0$, for which the density matrix $\rho$ can be neglected. We introduce a orthonormal basis 
  \begin{equation}\label{eqn:eachsiteoperator}
  \{\hat{O}_j^m\}=\{I,\hat c_j+\hat c_j^\dagger,i(\hat c_j-\hat c_j^\dagger),2\hat c_j^\dagger \hat c_j-1\},
  \end{equation}
  with $m\in\{1,2,3,4\}$ on each system fermion mode \cite{Zhang:2023vpm}. Here, the addition of $-1$ to the particle number operator ensures orthogonality with the identity. We assign a size to each local basis operator as $\{s_m\}=\{0,1,1,2\}$. This definition matches the standard convention \cite{roberts2018operator} if we split each complex fermion into two Majorana fermions. A generic operator $\hat{O}(t)$ can be expanded as 
  \begin{equation}
   \hat{O}(t)=\sum_{\{m_i\}}C_{m_1m_2...m_N}(t)\hat O_1^{m_1}\hat O_2^{m_2}...\hat O_N^{m_N}\otimes \hat{O}_E,
  \end{equation}
  where $\hat{O}_E$ is an operator supported solely within the environment. We assume $\hat{O}$ is some simple operator, such as $\hat{c}_1$. We define the size of each string operator $\hat O_1^{m_1}...\hat O_N^{m_N}\otimes \hat{O}_E$ as a summation $s_{\{m_i\}}=\sum_{i=1}^N s_{m_i}$. The operator size distribution $P_0(n,t)$ at half-filling $\mu=0$ then reads
  \begin{equation}
  P_0(s,t)=\sum_{\{m_i\}}\left|C_{m_1m_2...m_N}(t)\right|^2\delta_{s,s_{\{m_i\}}} \langle \hat{O}_E^\dagger \hat{O}_E\rangle.
  \end{equation}
  Here, $\langle \hat{O}\rangle\equiv 2^{-M-N}\text{tr}[\hat{O}]$ measures the norm of the environment operator. For later convenience, we further introduce its generating function $Z_0(\nu,t)=\sum_s e^{-\nu s} P_0(s,t)$. 

  By generalizing analysis in closed systems \cite{Zhang:2023vpm}, we can express $Z(\mu,t)$ as a correlation function on the doubled Hilbert space by introducing an auxiliary copy of the total system, which includes system fermions $\hat \xi_j$ and environment fermions $\eta_a$. We pair up $(c_j,\xi_j)$ and $(e_a,\eta_a)$ by introducing a maximally entangled state 
  \begin{equation}
  |I\rangle=\otimes_{j=1}^N\left[\frac{|01\rangle_j+|10\rangle_j}{\sqrt{2}}\right]\otimes_{a=1}^M\left[\frac{|01\rangle_a+|10\rangle_a}{\sqrt{2}}\right].
  \end{equation}
  With the help of $|I\rangle$, we can map each operator $\hat{O}(t)$ into a state $|\hat O(t)\rangle=\hat{O}(t) |I\rangle$. We then look for a size operator $\hat{S}_0=\sum_{j=1}^N \hat s_{0,j}$, which satisfies  
  \begin{equation}
  \hat s_{0,j}|\hat{O}^m_j\rangle=s_m |\hat{O}^m_j\rangle.
  \end{equation}
  It is straightforward to check that we can choose
  \begin{equation}
  \hat s_{0,j}=\frac{1}{2}\left[(c_j^\dagger-\xi_j^\dagger)(c_j^{}-\xi_j^{})+(c_j^{}+\xi_j^{})(c_j^\dagger+\xi_j^{\dagger})\right].
  \end{equation}
  Therefore, the generating function can be expressed as 
  \begin{equation}
  Z_0(\nu,t)=\langle \hat{O}(t)|e^{-\nu \hat{S}_0}|\hat{O}(t)\rangle.
  \end{equation}

  Now, we consider adding a restriction to the Hilbert space to focus on the subspace with a general density $n$. The situation here closely parallels the definition of operator size at finite temperature \cite{gu2022two,Zhang:2022fma}. In the grand canonical ensemble, the restriction can be incorporated by replacing $|\hat{O}(t)\rangle$ with
  \begin{equation}
  |\hat{O}(t)\rangle_\mu\equiv\frac{|\rho^{\frac{1}{4}}\hat{O}(t)\rho^{\frac{1}{4}}\rangle }{\sqrt{\langle \rho^{\frac{1}{4}}\hat{O}(t)\rho^{\frac{1}{4}}|\rho^{\frac{1}{4}}\hat{O}(t)\rho^{\frac{1}{4}}\rangle}}=\frac{|\rho^{\frac{1}{4}}\hat{O}(t)\rho^{\frac{1}{4}}\rangle}{\sqrt{\langle\hat{O}\rho^{\frac{1}{2}}\hat{O}\rho^{\frac{1}{2}}\rangle}}.
  \end{equation}
  Then, we choose to modify the size operator $\hat{S}_\mu\equiv \sum_{j=1}^N\hat{s}_{\mu,j}$ correspondingly such that identity operator has no size $\hat{s}_{\mu,j}|I\rangle_\mu=0$. A convenient choice is \cite{Zhang:2023vpm} 
  \begin{equation}\label{eqn:size}
  \begin{aligned}
  \hat{s}_{\mu,j}={A(\mu)}\Big[&(e^{\frac{\mu}{4}}\hat c_j^\dagger-e^{-\frac{\mu}{4}}\hat \xi_j^\dagger)(e^{\frac{\mu}{4}}\hat c_j^{}-e^{-\frac{\mu}{4}}\hat \xi_j^{})\\
  &+(e^{-\frac{\mu}{4}}\hat c_j^{}+e^{\frac{\mu}{4}}\hat \xi_j^{})(e^{-\frac{\mu}{4}}\hat c_j^\dagger+e^{\frac{\mu}{4}}\hat \xi_j^{\dagger})\Big].
  \end{aligned}
  \end{equation}
  Here, $A(\mu)$ is an overall normalization factor that should satisfy $A(0)=1/2$. The size operator $\hat{s}_{\mu,j}$ is positive semi-definite and measures "excitations" relative to the reference state $|I\rangle_\mu$. In this work, we fix $A(\mu)=({2\cosh (\mu/2)})^{-1}=\sqrt{n(1-n)}$ by further requiring 
  \begin{equation}
  \hat{s}_{\mu,j}|\hat{c}_j\rangle_\mu=|\hat{c}_j\rangle_\mu.
  \end{equation}
  Putting all ingredients together, we define the generating function at finite chemical potential as 
  \begin{equation}\label{eqn:sizezz}
  Z_\mu(\nu,t)=~_\mu\langle \hat{O}(t)|e^{-\nu \hat{S}_\mu}|\hat{O}(t)\rangle_\mu,
  \end{equation}
  and the operator size distribution $P_\mu(s,t)$ is its inverse Laplace transform. By definition, we have $Z_\mu(0,t)=1$, which fixes the normalization $\sum_s P_\mu(s,t)=1$.

\section{Evolution of Operator Size}
The information scrambling in chaotic systems with all-to-all interactions can be divided into two regimes: the early-time regime with $t\sim O(1)$ and the late-time regime $t\sim\ln N$. In Ref. \cite{Zhang:2023xrr}, we demonstrate the existence of the environment-induced information scrambling transition in Majorana systems by computing the size distribution for single fermion operator in the early-time regime. In this section, we extend the calculation to our complex Brownian SYK models with $\hat{O}=\hat{c}_1$. 

The study of $ Z_\mu(\nu,t)$ requires introducing the doubled Keldysh contour \cite{aleiner2016microscopic}. As a warm-up, we first consider the average operator size $\overline{S}$, which reads
\begin{equation}
\begin{aligned}
\overline{S}=&~_\mu\langle \hat{c}_1(t)|\hat{S}_\mu|\hat{c}_1(t)\rangle_\mu= \sum_j\langle \rho^{\frac{1}{2}}\{\hat{c}_1(t),\hat{c}_j\} \rho^{\frac{1}{2}}\{\hat{c}_1(t),\hat{c}_j\} \rangle\\&+\sum_j\langle \rho^{\frac{1}{2}}\{\hat{c}_1(t),\hat{c}_j^\dagger\} \rho^{\frac{1}{2}}\{\hat{c}_1(t),\hat{c}_j^\dagger\} \rangle,
\end{aligned}
\end{equation}
where, we have used $\langle\hat{c}_1^\dagger\rho^{\frac{1}{2}}\hat{c}_1\rho^{\frac{1}{2}}\rangle=\sqrt{n(1-n)}$. In the path-integral approach, these four-point functions correspond to
\begin{equation}\label{eqn:doubleKeldysh}
\begin{tikzpicture}[scale=0.65,baseline={([yshift=0pt]current bounding box.center)}]

\draw[thick,blue,far arrow] (0pt,0pt)--(140pt,0pt);
\draw[thick,blue,far arrow] (140pt,-4pt)--(0pt,-4pt);
\filldraw (140pt,-2pt) circle (2pt) node[above right]{ };
\node at (70pt,6pt) {\scriptsize $u$};
\node at (70pt,-10pt) {\scriptsize $d$};
\filldraw (0pt,-2pt) circle (0pt) node[above left]{ };
\draw[thick,blue] (0pt,0pt)--(0pt,10pt);
\draw[thick,blue] (-2pt,7pt)--(2pt,7pt);
\draw[thick,blue] (0pt,-4pt)--(0pt,-34pt);
\node[svertex] at (0pt,-19pt) {\tiny$\sqrt{\rho}$};
\draw[thick,blue,far arrow] (0pt,-34pt)--(140pt,-34pt);
\draw[thick,blue,far arrow] (140pt,-38pt)--(0pt,-38pt);
\filldraw (140pt,-36pt) circle (2pt) node[above right]{};
\node at (70pt,-28pt) {\scriptsize $u$};
\node at (70pt,-44pt) {\scriptsize $d$};
\draw[thick,blue] (0pt,-38pt)--(0pt,-72pt);
\node[svertex] at (0pt,-53pt) {\tiny$\sqrt{\rho}$};
\draw[thick,blue] (-2pt,-69pt)--(2pt,-69pt);
\node at (170pt,-2pt) {\scriptsize world $1$};
\node at (170pt,-36pt) {\scriptsize world $2$};
\filldraw[red] (20pt,0pt) circle (1.5pt);
\filldraw[red] (20pt,-4pt) circle (1.5pt);
\filldraw[red] (20pt,-34pt) circle (1.5pt);
\filldraw[red] (20pt,-38pt) circle (1.5pt);
\end{tikzpicture}
\end{equation}
In addition to the index $s\in\{u,d\}$, we further introduce a world index $w\in\{1,2\}$ that distinguishes the two forward/backward evolution branches. Here, black dots represent the insertion of $\hat{c}_1$ or $\hat{c}_1^\dagger$. Red dots represent the source term due to the size operator $\hat{S}_\mu$. In other words, we have 
\begin{equation}
\begin{aligned}
\overline{S}=&\sum_j\left<T_\mathcal C \,c^{u,2}_1(t)\overline{c}^{u,1}_1(t)
  \left[c_j^{u,2}(0)-c_j^{d,2}(0)\right]\left[\bar{c}_j^{d,1}(0)-\bar{c}_j^{u,1}(0)\right]\right>\\
  &+\left<T_\mathcal C \,c^{u,2}_1(t)\overline{c}^{u,1}_1(t)
  \left[\bar{c}_j^{u,2}(0)-\bar{c}_j^{d,2}(0)\right]\left[c_j^{d,1}(0)-c_j^{u,1}(0)\right]\right>.
  \end{aligned}
\end{equation}
Here, the minus sign arises from the anti-commutation relation of Grassmann fields, and $T_\mathcal C$ denotes the contour ordering operator. Generalizing the discussion to the generating function $Z_\mu(\nu,t)$, we have 
\begin{equation}
\begin{aligned}
Z_\mu(\nu,t)&=\frac{\left<T_\mathcal C\,c^{u,2}_1(t)\overline{c}^{u,1}_1(t)e^{-\Delta I}\right>}{\sqrt{n(1-n)}},
  \end{aligned}
\end{equation}
with souce term 
\begin{equation}
\begin{aligned}
{\Delta I}=C(\nu,\mu)
  &\sum_j\Bigg(\left[c_j^{u,2}(0)-c_j^{d,2}(0)\right]\left[\bar{c}_j^{d,1}(0)-\bar{c}_j^{u,1}(0)\right]\\
&-
  \left[\bar{c}_j^{u,2}(0)-\bar{c}_j^{d,2}(0)\right]\left[c_j^{d,1}(0)-c_j^{u,1}(0)\right]\Bigg),
\end{aligned}
\end{equation}
where $C(\nu,\mu)=(1-e^{-\nu})A(\mu)$ denotes the strength of the source term. 

To derive the equation for $Z_\mu(\nu,t)$, we introduce the interworld Wightmann function \begin{equation}
G^W(t_1,t_2)=\left<T_\mathcal C\,c^{u,2}(t_1)\bar{c}^{u,1}(t_2)e^{-\Delta I}\right>,
\end{equation}
Using the symmetry when interchanging two worlds, this definition can be equivalently expressed as $G^W(t_1,t_2)=\left<T_\mathcal C\,\bar{c}^{u,2}(t_2)c^{u,1}(t_1)e^{-\Delta I}\right>$. The interworld Wightmann function satisfies the Schwinger-Dyson equation \cite{Zhang:2020jhn,Zhang:2023xrr}
\begin{equation}\label{eqn:SDGW}
G^W(t_1,t_2)=\int dt_3dt_4~G^R(t_1-t_3) \Sigma^W(t_3,t_4)G^A(t_4-t_2). 
\end{equation}
with self-energy
\begin{equation}
\begin{aligned}
\Sigma^W(t,t')=&\delta(t-t')\Bigg(\sum_qJ_q G^W(t,t)^{q-1}- C(\nu,\mu)\delta(t)\\
&+{\sum_{\{p_l\}}}'U_{\{p_l\}}G^W(t,t)^{p_1+p_2-1}[n(1-n)]^{\frac{p_3+p_4}{2}}\Bigg).
\end{aligned}
\end{equation}
Finally, we set $t_1=t_2=t$ and apply the operator $(\partial_t+\Gamma)$ to both sides of \eqref{eqn:SDGW}. Using the relation $G^W(t,t)=\sqrt{n(1-n)}Z_\mu(\nu,t)$, we find the differential equation for $t>0$
\begin{equation}\label{eqn:Zmu}
\begin{aligned}
\frac{\partial Z_\mu}{\partial t}=&\sum_qJ_q [n(1-n)]^{\frac{q}{2}-1}(Z_\mu^{q-1}-Z_\mu)\\&+{\sum_{\{p_l\}}}' U_{\{p_l\}} [n(1-n)]^{\frac{\sum_l p_l}{2}-1}(Z_\mu^{p_1+p_2-1}-Z_\mu),
\end{aligned}
\end{equation} 
with initial condition $Z_\mu(0,\nu)=e^{-\nu}$ fixed by the source term $C(\nu,\mu)$.

\section{Information Scrambling Transition}
The terms on the R.H.S. of \eqref{eqn:Zmu} follow the same pattern as their counterparts for Majorana fermions \cite{Zhang:2023xrr}. Therefore, we expect the system to exhibit an environment-induced information scrambling transition, as we will now elaborate. We first study the average size $\overline{S}(t)=-\partial_\nu Z_\mu(t,0)$. Using \eqref{eqn:Zmu}, it is straightforward to derive $\overline{S}'=\varkappa \overline{S}$, with
\begin{equation}\label{eqn:St}
\begin{aligned}
\varkappa=&\sum_qJ_q [n(1-n)]^{\frac{q}{2}-1}(q-2)\\&+{\sum_{\{p_l\}}}' U_{\{p_l\}} [n(1-n)]^{\frac{\sum_l p_l}{2}-1}(p_1+p_2-2).
\end{aligned}
\end{equation}
Therefore, terms can be divided into three groups:
\begin{enumerate}
\item Terms with $q=2$ or $p_1+p_2=2$ vanish, as they do not contribute to the operator growth.

\item Terms with $p_1+p_2=1$ are negative, representing the dissipation that suppreses the operator growth.

\item Terms with $q\geq 4$ or $p_1+p_2>1$ are positive, representing the system intrinsic or environment propelled scramblings respectively. 
\end{enumerate}
When the dissipation effect dominates, we find $\varkappa<0$, and the operator size decays exponentially in time. When scrambling dominates, we have $\varkappa>0$, and the operator size grows exponentially in the early-time regime. These two dynamical phases are separated by the transition point at $\varkappa=0$, where the average size remains constant. It is also interesting to note that the dissipation effect only manifests when the U(1) symmetry is weak, as terms with $p_1+p_2=1$ do not conserve particle numbers in the system. Therefore, systems with a strong U(1) symmetry shall always lie in the scrambling phase.

      \begin{figure}[tb]
    \centering
    \includegraphics[width=0.99\linewidth]{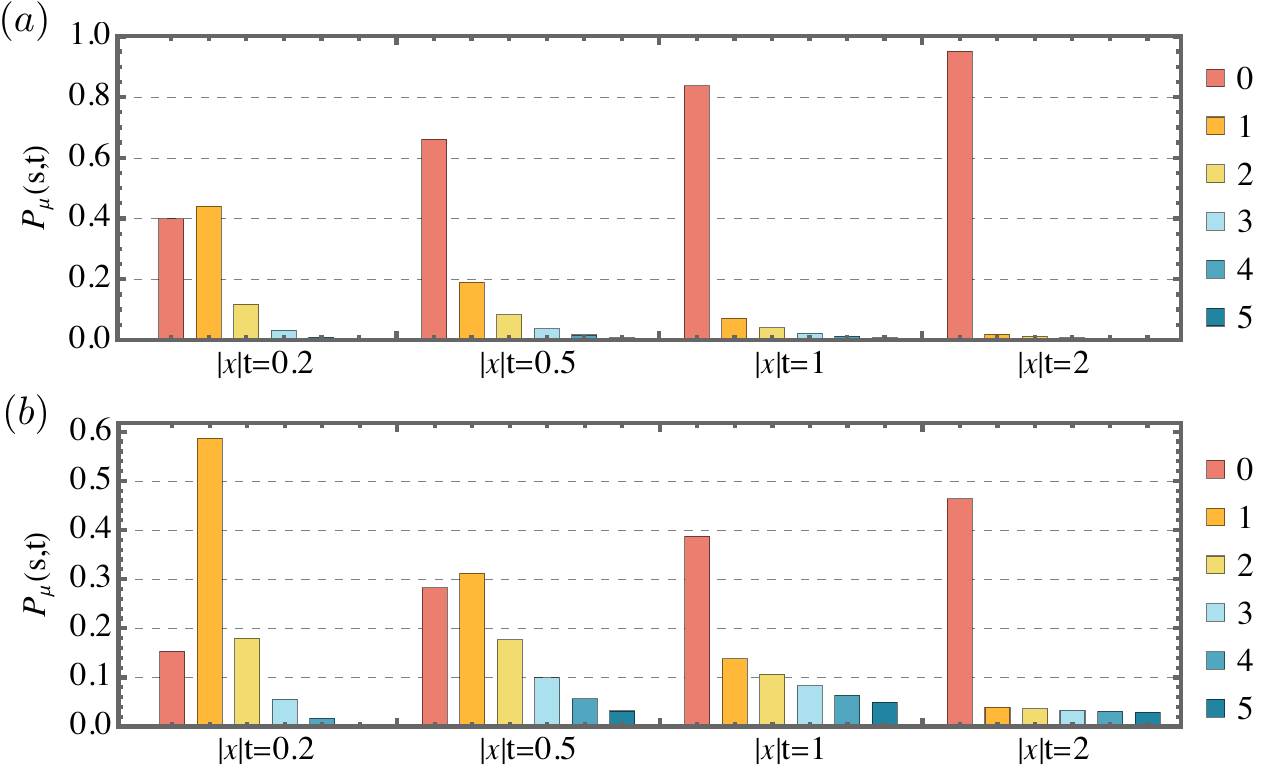}
    \caption{The early-time operator size distribution $P_\mu(s,t)$ for the model (\ref{hse}) with only $U_1$ and $U_3$ nonzero across different phases: (a) In the dissipative phase with $U_1=3U_3[n(1-n)]/2$, the operator weight rapidly decays to zero for $s\geq 1$. (b) In the scrambling phase with $U_1=U_3[n(1-n)]/2$, the operator weight at $s=0$ saturates to $P_\mu(0,\infty)=r=1/2$. }
    \label{fig:Early-time_size}
  \end{figure}

We further explore the fate of the scrambling transition in the low/high-density limit $n\rightarrow 0/1$. In this case, \eqref{eqn:St} is dominated by terms with the smallest $q$ or $\sum_l p$. For the dissipation process, we only keep $p_1+p_2=1$ and $p_3+p_4=1$ with strength $U_{1,0,0,1}\equiv U_1$. This represents a direct hopping between the system and the environment. The dominant scrambling process is given by $q=4$ or $\sum_lp_l=4$, which are four-fermion interactions. As an example, let us focus on a special term with $p_1+p_2=3$ and introduce its strength $U_3$. This corresponds to the interaction between three system fermion operators and one environment fermion operator. This environment propelled scrambling rate is proportional to $n(1-n)$ due to the Pauli blockade in the scattering process. This model is illustrated in FIG. \ref{fig:schemticas}, for which we have $\varkappa=U_3[n(1-n)]-U_1$, and the scrambling transition occurs at $U_1/U_3 = n(1-n)$. We expect its scaling at $n\rightarrow 0/1$ to be exact in wider generic models. 

To fully characterize the operator growth in this simplified model with only $U_1$ and $U_3$, we directly solve the differential equation \eqref{eqn:Zmu}. The result reads
\begin{equation}\label{eqn:solz}
    Z_\mu(\nu,t)=1-\frac{(r-1)(1-e^{-\nu})}{e^{-\varkappa t}(r-e^{-\nu})-(1-e^{-\nu})}.
   \end{equation}
  This takes the same form as the results of the Majorana model in \cite{Zhang:2023xrr}. Here, all density dependence is through $\varkappa$ and $r \equiv U_1/U_3[n(1-n)]$. Expanding in terms of $e^{-\nu}$, we obtain the operator size distribution  
   \begin{equation}\label{eqn:solP}
   P_\mu(s,t)=\begin{cases}
      1-(1-r)\frac{e^{\varkappa t}}{e^{\varkappa t}-r}, & (s=0)\\
      (1-r)^2\frac{e^{\varkappa t}(e^{\varkappa t}-1)^{s-1}}{(e^{\varkappa t}-r)^{s+1}}. & (s\geq 1)
    \end{cases} 
   \end{equation}
The dissipative phase corresponds to $r>1$ and $\varkappa<0$, where the operator size distribution decays exponentially in size $s$. For $|\varkappa| t \gtrsim 1$, the operator weight at $s>0$ also decays as $e^{\varkappa t}$ in time, approaching $P_\mu(s,\infty)=\delta_{s,0}$ in the long-time limit. As a comparison, the scrambling phase has $r<1$ and $\varkappa>0$. The operator size distribution is approximately constant for $1 \leq s \lesssim e^{\varkappa t}$. In the long-time limit, we have $P_\mu(0,\infty)=r$, and there is always finite operator weight at $s>0$. We provide a density plot of $P_\mu(s,t)$ in both phases in FIG. \ref{fig:Early-time_size}.

\section{Scrambling Phase at Late-time}
In the scrambling phase, the growth of operator size saturates when the time $t$ greatly exceeds the scrambling time $t_s=\varkappa^{-1}\ln N$. In this section, we examine the operator size distribution in the late-time regime when $t\approx t_s$. Since a typical operator now has size $s\sim O(N)$, we take the continuum limit by introducing $\sigma =s/N \in[0,1]$ and $\mathcal{P}_\mu(\sigma,t)=NP_\mu(\sigma N, t)$ as in Majorana fermion systems \cite{Zhang:2023xrr}. As in the early-time calculation, it is convenient to introduce the generating function
\begin{equation}
\mathcal{S}_\mu(v,t)=\int_0^1 d\sigma~\mathcal{P}_\mu(\sigma,t)e^{-\sigma v}=Z_\mu(v/N,t).
\end{equation}

Due to the exponential growth factor $e^{\varkappa t}$ approaching order $N$, the traditional large-$N$ expansion becomes problematic. Consequently, Eq. \eqref{eqn:SDGW} or \eqref{eqn:Zmu} becomes invalid. To perform a resummation over all leading contributions for fixed $e^{\varkappa t}/N$, we employ the scramblon effective field developed in Refs.~\cite{gu2022two,Stanford:2021bhl,Zhang:2022fma}. The basic assumption is that these contributions arise due to out-of-time-ordered correlations mediated by a collective mode, known as the scramblon. This effective theory contains a set of Feynman rules for scramblons, which can be used to compute generic correlation functions. Firstly, a scramblon has an exponentially growing propagator
\begin{equation}
\begin{tikzpicture}[scale=1.5,baseline={([yshift=-3pt]current bounding box.center)}]

\draw[thick,wavy] (0pt,0pt) to (30pt,0pt);

\filldraw  (0pt,0pt) circle (0pt) node[left]{\scriptsize $t$};
\filldraw  (30pt,0pt) circle (0pt) node[right]{\scriptsize $0$};
\end{tikzpicture}= -\frac{e^{\varkappa t}}{C}\equiv -\lambda,
\end{equation}
which is a signature of quantum many-body chaos. Here, $C\propto N$ is a numerical factor. Secondly, scramblons can interact with a pair of fermion operators $\hat{c}_j$ and $\hat{c}_j^\dagger$, with scattering vertexes  
\begin{equation}
\begin{tikzpicture}
\node[bvertexnormal] (R) at (-30pt,0pt) {};
\draw[thick] (R) -- ++(135:16pt) node[left]{};
\draw[thick] (R) -- ++(-135:16pt) node[left]{};
\draw[wavy] (R) -- ++(45:16pt) node[left]{};
\draw[wavy] (R) -- ++(-45:16pt) node[left]{};
\draw[wavy] (R) -- ++(0:16pt) node[left]{};
\end{tikzpicture}=\Upsilon^{R,m},\ \ \ \ \ \ 
\begin{tikzpicture}
\node[bvertexnormal] (A) at (30pt,0pt) {};
\draw[thick] (A) -- ++(45:16pt) node[right]{};
\draw[thick] (A) -- ++(-45:16pt) node[right]{};
\draw[wavy] (A) -- ++(135:16pt) node[left]{};
\draw[wavy] (A) -- ++(-135:16pt) node[left]{};
\draw[wavy] (A) -- ++(180:16pt) node[left]{};
\end{tikzpicture}=\Upsilon^{A,m}.
\end{equation} 
Here, $m$ labels the number of scramblons, and $R/A$ denotes scattering in the past/future. Notice that we have not fixed the normalization of the scramblon field. There exists a transformation $C \rightarrow a^2C$ and $\Upsilon^m\rightarrow a^m \Upsilon^m$ that leaves all physical observables invariant. For systems with time-reversal symmetry (as our complex Brownian SYK model), we have $\Upsilon^{R,m}=\Upsilon^{A,m}=\Upsilon^{m}$. In particular, we have $\Upsilon^0=G^W(0,0)|_{\nu=0}=\sqrt{n(1-n)}$.

Let us first compute the generating function $\mathcal{S}_\mu(v,t)$ using these Feynman rules and postpone the determination of $C$ and $\Upsilon^{m}$. If we only keep out-of-time order correlations \cite{gu2022two}, we can approximate 
\begin{equation}
\mathcal{S}_\mu(v,t)\approx ~_\mu\langle \hat{c}_1(t)|e^{-\frac{v}{N} \hat{S}_\mu'}|\hat{c}_1(t)\rangle_\mu,
\end{equation}
with 
\begin{equation}\label{eqn:size}
  \begin{aligned}
  \hat{S}_\mu'=\sum_j{2A(\mu)}\Big[&2\sqrt{n(1-n)}-\hat{\xi}_j^\dagger \hat{c}_j-\hat{c}_j^\dagger\hat{\xi}_j \Big].
  \end{aligned}
  \end{equation}
Here, we retain cross terms between $(\hat{c},\hat{c}^\dagger)$ and $(\hat{\xi},\hat{\xi}^\dagger)$ as operators while replacing other terms with their expectation values on the reference state $|I\rangle_\mu$. We further represent $\mathcal{S}_\mu(v,t)$ using the path-integral approach:
\begin{equation}
\mathcal{S}_\mu=e^{-4 A \sqrt{n(1-n)} v}\frac{\left<T_\mathcal C\,c^{u,2}_1(t)\overline{c}^{u,1}_1(t)e^{\frac{2A v}{N}\sum_j (\,c^{d,2}_1(0)\overline{c}^{d,1}_1(0)+\bar c^{d,2}_1(0)c^{d,1}_1(0))}\right>}{\sqrt{n(1-n)}},
\end{equation}
We expand the source term in $\nu$, each insertion can create arbitrary number of scramblons, which interacts with $c^{u,2}_1(t)\overline{c}^{u,1}_1(t)$ in the future. 
Diagrammatically, we find 
\begin{equation}\label{eqn:summation}
\begin{aligned}
\mathcal{S}_\mu(v,t)=&
\begin{tikzpicture}
\node[bvertex] (R) at (-0pt,0pt) {};
\node[bvertexsmall] (A1) at (30/1.4142pt,30/1.4142pt) {};
\node[bvertexsmall] (A2) at (30pt,0pt) {};
\node[bvertexsmall] (A3) at (30/1.4142pt,-30/1.4142pt) {};

\draw[thick] (R) -- ++(135:26pt) node[left]{\scriptsize$ $};
\draw[thick] (R) -- ++(-135:26pt) node[left]{\scriptsize$ $};
\draw[thick] (A1) -- ++(45+45:10pt) node{};
\draw[thick] (A1) -- ++(-45+45:10pt) node{};
\draw[thick] (A2) -- ++(45:10pt) node{};
\draw[thick] (A2) -- ++(-45:10pt) node{};
\draw[thick] (A3) -- ++(45-45:10pt) node{};
\draw[thick] (A3) -- ++(-45-45:10pt) node{};

\draw[wavy] (R) to[out=60,in=-150] (A1);
\draw[wavy] (R) to[out=30,in=-120] (A1);
\draw[wavy] (R) to[out=15,in=180-15] (A2);
\draw[wavy] (R) to[out=-15,in=180+15] (A2);
\draw[wavy] (R) to (A3);
\end{tikzpicture}
=\frac{e^{-4 A \sqrt{n(1-n)} v}}{\sqrt{n(1-n)}}\sum_{n=0}^{\infty} \frac{(4A v)^n}{n!}\\&\times\sum_{\{m_j\geq 0\}} \left(-\frac{e^{\varkappa t}}{C}\right)^{ \sum_jm_j}\frac{  \Upsilon^{\sum_j m_j} \Upsilon^{m_1}\ldots \Upsilon^{m_n} }{m_1! m_2! \ldots m_n! },
\end{aligned}
\end{equation}
where we have used the fact that both $c^{d,2}_1(0)\overline{c}^{d,1}_1(0)$ and $\bar c^{d,2}_1(0)c^{d,1}_1(0)$ share the same scattering vertexes \cite{Zhang:2023vpm} in our setup where the density matrix is equally split as shown in \eqref{eqn:doubleKeldysh}. In other words, $\Upsilon^m$ remains invariant under the particle-hole transformation. The summation can be carried out by introducing auxiliary functions $f(x)$ and $h(y)$ \cite{gu2022two}, which satisfies
\begin{equation}
f(x)=\sum_{m=0}^\infty \frac{(-x)^m}{m!}\Upsilon^m,\ \ \ \ \ \ f(x)=\int_0^\infty dy~h(y)e^{-xy}.
\end{equation}
In particular, we have $f(0)=\Upsilon^0=\sqrt{n(1-n)}$. The summation in \eqref{eqn:summation} gives
\begin{equation}\label{eqn:resSmunut}
\mathcal{S}_\mu(v,t)=\int_0^\infty dy~\frac{h(y)}{\sqrt{n(1-n)}} e^{-4A v\left(\sqrt{n(1-n)}-f( \lambda y)\right)}.
\end{equation}

Now, we determine $\Upsilon^m$ in the simplified model by comparing \eqref{eqn:resSmunut} with \eqref{eqn:solz} in the early-time limit with $1 \ll \varkappa t \ll \ln N$. Using $\eqref{eqn:resSmunut}$, we should expand $e^{\varkappa t} y/C$ to leading order, which gives
\begin{equation}
\mathcal{S}_\mu(v,t)=\frac{1}{\sqrt{n(1-n)}}f\left(\frac{4A(\mu) e^{\varkappa t}\Upsilon^1 v}{C}\right).
\end{equation}
On the other hand, expanding \eqref{eqn:solz} for small $\nu=v/N$, we obtain
\begin{equation}
\mathcal{S}_\mu(v,t)=r+\frac{1-r}{1+e^{\varkappa t}v(1-r)^{-1}/N}.
\end{equation}
To proceed, we use the aforementioned symmetry $C \rightarrow a^2C$ and $\Upsilon^m\rightarrow a^m \Upsilon^m$ to fix 
\begin{equation}\label{eqn:consistency}
\frac{4A(\mu) \Upsilon^1 }{C}=\frac{1}{(1-r)N}.
\end{equation}
In this convention, we have $f(x)=\sqrt{n(1-n)}[r+(1-r)/(1+x)]$, which corresponds to
\begin{equation}
\Upsilon^0=\sqrt{n(1-n)},\;\;\;\;\; \Upsilon^m=\sqrt{n(1-n)}(1-r)\,m!.
\end{equation}
Using \eqref{eqn:consistency}, this further gives $C/N=4(1-r)n(1-n)$. The scaling of $\Upsilon^m$ and $C$ in the low/high density limit matches their counterparts in closed Brownian SYK models \cite{Zhang:2023vpm}. We can also apply the inverse Laplace transform to $f(x)$ to determine
\begin{equation}
h(y)=\sqrt{n(1-n)}\left[(1-r)e^{-y}+r\delta(y)\right].
\end{equation}

Using explicit expressions for both $f(x)$ and $h(y)$, the generating function reads
\begin{equation}\label{eqn:resSmunut2}
\mathcal{S}_\mu(v,t)=r+\int_0^\infty dy~(1-r)e^{-y} e^{-s_\text{sc}\frac{\lambda y}{1+\lambda y}},
\end{equation}
where $s_\text{sc}=4n(1-n)(1-r)$. Performing the inverse Laplace transform, we find the continuum size distribution function $\mathcal{P}_\mu(\sigma,t)=r\delta(\sigma)+\mathcal{P}^{\text{r}}_\mu(\sigma,t)$, with
\begin{equation}
\begin{aligned}
\mathcal{P}^{\text{r}}_\mu(\sigma,t)&=\theta(s_\text{sc}-\sigma)\frac{s_\text{sc} (1-r) e^{\frac{\sigma }{\lambda  (\sigma-s_\text{sc})
   }}}{\lambda  (\sigma -s_\text{sc})^2}.
\end{aligned}
\end{equation}
For systems at half-filling $n=1/2$ with $r=0$, the operator size distribution saturates to a delta function at $\sigma =s_\text{sc}=1$. This corresponds to the standard maximally scrambled operator in complex fermion systems, where each operator in \eqref{eqn:eachsiteoperator} appears with the same probability, resulting in an average size of $1$. When $n\neq 1/2$, the Hilbert space restriction reduces the effective dimension of the operator space, and the saturation value becomes $4n(1-n)$, which diminishes as $n\rightarrow 0/1$. Introducing a finite $r$ further reduces $s_{\text{sc}}$ by a factor of $(1-r)$, indicating a ``drag" force exerted by the environment, similar to the Majorana case \cite{Zhang:2023xrr}. Additionally, we have a singular part $r\delta(\sigma)$, representing the contribution of pure environment operators.

\section{Summary}
In this work, we investigate how charge conservation affects the environment-induced information scrambling transitions by examining a solvable complex Brownian SYK model embedded in an environment. After defining the size operator within our setup, we analyze the dominant processes in the low/high density limits. This analysis leads us to identify a minimal model featuring system-environment hopping, denoted as $U_1$, and four-fermion interactions, denoted as $U_3$. We determine that the scrambling transition occurs at $U_1/U_3=n(1-n)$, which separates a dissipative phase characterized by large $U_1$ from a scrambling phase characterized by small $U_1$. Furthermore, we delve into the late-time regime using the scramblon effective theory \cite{gu2022two,Stanford:2021bhl}, revealing the renormalization of maximally scrambled operators with a typical size of $4n(1-n)(1-r)$.

\vspace{5pt}
\textit{Acknowledgment.} This project is supported by the Key Area Research and Development Program of Guangdong Province (Grant No. 2019B030330001), the National Natural Science Foundation of China (Grant No.~12374477 and No.~12074440), and Guangdong Project (Grant No.~2017GC010613).

\bibliography{ref.bib}

\end{document}